\documentclass[aps,reprint,article,prx]{revtex4-2}
\usepackage{amsmath}
\usepackage{dcolumn}
\usepackage{graphicx}
\usepackage{float}
\usepackage{color}
\usepackage{booktabs}
\usepackage{multirow}

\begin{document}

\title{Diminishing Mott gap by doping electrons through depositing one monolayer thin film of Rb on Ca$_{2}$CuO$_{2}$Cl$_{2}$}

\author{Han Li, Zhaohui Wang, Shengtai Fan, Huazhou Li, Huan Yang,$^*$ and Hai-Hu Wen$^\dag$}

\affiliation{National Laboratory of Solid State Microstructures and Department of Physics, Collaborative Innovation Center of Advanced Microstructures, Nanjing University, Nanjing 210093, China}

\begin{abstract}
Understanding the doping evolution from a Mott insulator to a superconductor probably holds the key for resolving the mystery of unconventional superconductivity in copper oxides. To elucidate the evolution of the electronic state starting from the Mott insulator, we dose the surface of the parent phase Ca$_{2}$CuO$_{2}$Cl$_{2}$ by depositing one monolayer thin film of Rb atoms which are supposed to donate electrons to the CuO$_{2}$ planes underneath. We successfully achieved the Rb thin films with periodic structures, and the scanning tunneling microscopy or spectroscopy (STM or STS) measurements on the surface show that the Fermi energy is pinned within the Mott gap but more close to the edge of the charge transfer band (CTB). However, the electron doping does not reduce the spectra weight of the upper Hubbard band (UHB) for the double occupancy as expected from the rigid model, but instead increase it; meanwhile, further doping will create a new wide spread in gap states derivative from the UHB, and the Mott gap will be significantly diminished. Our results provide new clues to understand the strong correlation effect of parent Mott insulators for cuprates and shed new light on the origin of high-temperature superconductivity.
\end{abstract}

\maketitle
\begin{figure*}
\includegraphics[width=18cm]{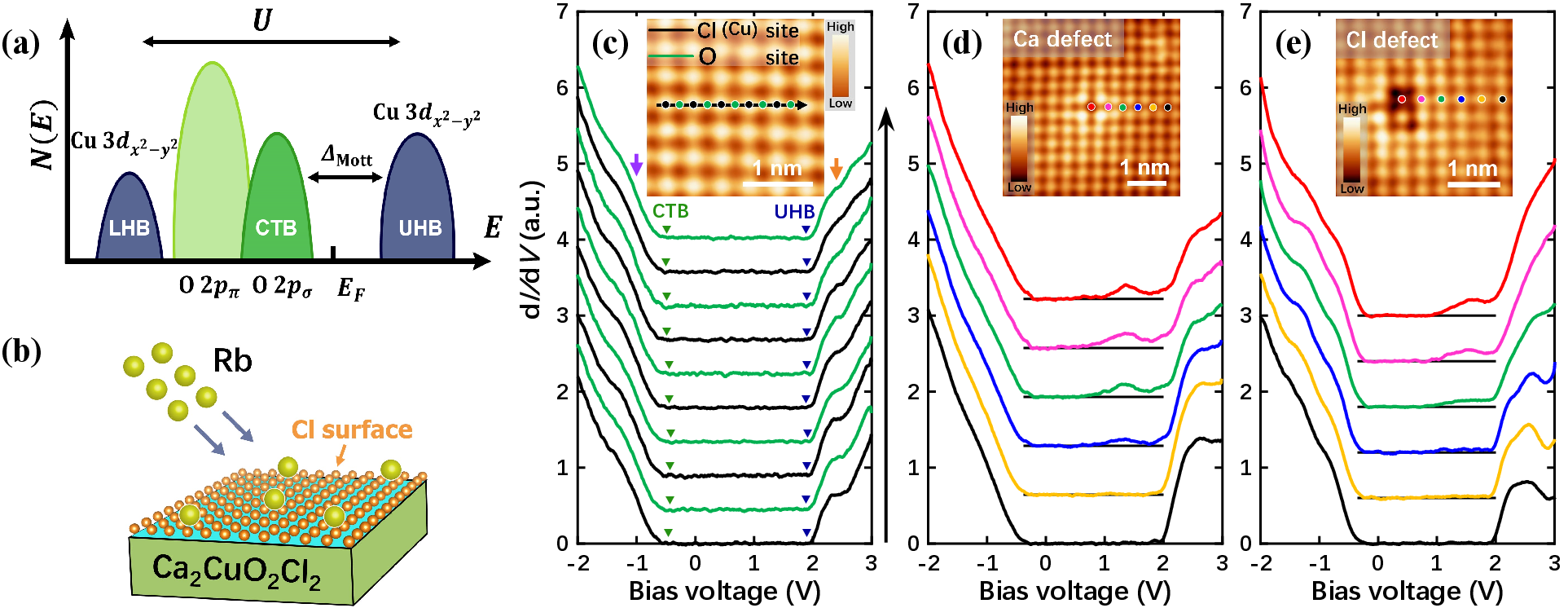}
\caption{(a) Schematic band structure of the parent Mott insulating phase of cuprates. (b) Schematic illustration of in situ Rb dosing onto CCOC surface. Electrons are introduced into the sample surface due to the physical absorption of Rb atoms. (c) Spatially resolved tunneling spectra measured along an arrow in the topography of Ca$_{2}$CuO$_{2}$Cl$_{2}$ shown as the inset ($V_\mathrm{set} = -2.5$ V, $I_\mathrm{set} = 100$ pA). The spectra in black are measured on Cl atoms on the surface or Cu atoms underneath, while the spectra in green are measured in the middle between two neighboring Cl atoms. Triangles in green mark tops of CTB, and those in blue mark bottoms of UHB. Purple and orange arrows roughly indicate the peak position of CTB and UHB, respectively. (d) A series of tunneling spectra taken at and away from a Ca defect ($V_\mathrm{set} = -2.5$ V, $I_\mathrm{set} = 100$ pA). (e) A series of tunneling spectra taken at and away from a Cl defect ($V_\mathrm{set} = -2.0$ V, $I_\mathrm{set} = 100$ pA). Tunneling spectra in all panels are shifted for clarity. Each spectrum in (d) and (e) is plotted using the same color as the point marking the measured position.} \label{fig1}
\end{figure*}

High-temperature superconductivity (HTSC) has continued to be the focus of studies in modern condensed matter physics since the discovery of cuprates. Despite significant efforts from both theoretical and experimental aspects, the mechanism of HTSC remains ambiguous. The band is half-filled in the parent compound of cuprates, however, the parent compound is a Mott insulator instead of the metal because of the strong coupling effect \cite{Xiaogang Wen,Kivelson}. The physics can be appropriately described by the half-filled Hubbard model \cite{Anderson 1959,RBV 1987}, in which a strong local Coulomb interaction $U$ prevents electrons from doubly occupying any single Cu $d_{x^2-y^2}$ orbital, resulting in the splitting of electronic states near Fermi energy into the so-called upper (UHB) and lower (LHB) Hubbard bands \cite{J. E. Hirsch}, see Fig.~\ref{fig1}(a). If the problem is further considered in the strong-coupling limit by the $t-J$ model \cite{t-J Model}, the low-energy electronic excitations of the charge-transfer band (CTB) are predicted due to the planar $\sigma$ hybridization of the Cu $d_{x^2-y^2}$ orbital with O $2 p_{x,y}$ orbitals \cite{J. Zaanen1}. Then the energy gap of the parent compound is the charge transfer energy difference $\Delta_\mathrm{Mott}$ instead of the strong on-site Coulomb interaction $U$ \cite{p-d model}. In addition, based on this paradigm, a robust long-range antiferromagnetic (AF) order is stabilized in the Cu lattice coexisting with the intriguing electronic physics in the parent compound of cuprates \cite{AFM}.

In cuprates, HTSC emerges by the chemical doping to the parent compound of the Mott insulator. Besides the novel competing or coexisting orders with superconductivity \cite{Kivelson,Stripe,CDW}, the charge transfer physics of doped Mott insulators may be the key point for unraveling HTSCs. In the parent compound, when the CuO$_{2}$ planes are doped with charge carriers, the AF phase is suppressed together with the Mott gap filling and the appearance of the quasiparticle-like density of states (DOS) near the Fermi level. One compelling explanation is that the chemical potential shifts towards the CTB (or the UHB) band to accommodate the doped holes (or electrons). Meanwhile, the many-body effect causes a transfer of the DOS from the UHB or CTB band to fill in the Mott gap \cite{Sawatzky1,W. A. Groen}. The doping evolution of the electronic structure in cuprates has been studied in various systems \cite{Xiaogang Wen,Ref1A,Ref1B,Ref1C,Ref1D,Ref1E}. One must consider how the doped charge carriers destroy the Mott state and how the initial pairing is formed. Our recent work is concerned with this issue, revealing the emergence of the low-energy gap in extremely hole underdoped cuprate superconductors \cite{Haihu Wen}.

CCOC is a canonical half-filling AF Mott insulator, and there is a single CuO$_{2}$ plane in one structural unit cell. The substitution of a monovalent Na$^+$ for divalent Ca$^{2+}$ results in hole doping of Ca$_{2-x}$Na$_{x}$CuO$_{2}$Cl$_{2}$ (Na-CCOC), and the maximum critical temperature is about $28$ K \cite{Takagi1}. A checkerboard electronic crystal state is observed in Na-CCOC \cite{Davis1,Takagi2}, and then it can be explained by nodal quasiparticles and antinodal charge ordering in this system \cite{Zhixun Shen1}. These results help understand the HTSC in cuprates. Moreover, electron doping has been achieved utilizing surface deposition of alkali atoms on its double-layer sibling system Ca$_{3}$Cu$_{2}$O$_{4}$Cl$_{2}$ \cite{Xingjiang zhou1}.

Here, we report the studies on the electronic-structure evolution of the CCOC system by dosing electrons through depositing Rb on the surface of the single crystal. We investigate the effect of the isolated Rb adatom/cluster as well as the Rb monolayer film with ordered lattice at $78$ K by scanning tunneling microscopy or spectroscopy (STM/STS).  At the Rb adatom/cluster, the DOS of the UHB is enhanced while some in-gap states appear near the foot of the UHB. On the Rb lattice, the in-gap states are significantly enhanced and the Mott gap is reduced. Our findings reveal how the Mottness collapses due to increasing electron concentration, which helps to understand the strong correlation effect of the parent phase with Mott insulating behavior, and sheds new light on the origin of HTSC.


Single crystals of CCOC were grown by the flux method \cite{Takano}. The powder mixture of CaO and CuCl$_{2}$ with a $2:1$ molar ratio was put into an alumina crucible. Then the crucible was heated to $780$ $^\circ$C and held at this temperature for $24$ hours to get the CCOC polycrystal. After an intermediate grinding, the precursor was heated to $780$ $^\circ$C and held for $5$ hours, then heated to $930$ $^\circ$C and held for $10$ hours. The ramp rate of temperature during the heating processes were $1 ^\circ$C/min. Finally, the melt was slowly cooled to room temperature at a rate of $1$ $^\circ$C/min. High-quality single crystals were found with the size of millimeters. Single crystals were cleaved in situ at room temperature in ultrahigh-vacuum conditions up to $1.9\times10^{-10}$ torr. The dosing of Rb atoms and STM/STS are performed with a cryogenic variable temperature ultrahigh vacuum STM (USM-$1300$, Unisoku Co., Ltd.). The Rb adatom/cluster or film growth was also carried out on the cleaved surface of CCOC. The Rb source (SAES group) was heated to about $600$ $^\circ$C (or $640$ $^\circ$C), and the growth time for the Rb flux was several seconds (or about $1$ minute) to get the Rb adatoms/clusters (or the Rb film). After the deposition, the sample was immediately put back into the STM head through transferring in the ultra high vacuum atmosphere, which was kept at a low temperature of about $78$ K. The electrochemically etched tungsten tip was used in this study, and the calibration was carried out on the single crystalline Au ($111$) surface. The tunneling spectra were recorded by using a standard lock-in amplifier technique with an ac oscillation of $25$ meV (root-mean-square amplitude) and $973$ Hz. All the tunneling data were taken at a temperature of about $78$ K. We have tried the STM/STS measurements at liquid He temperature, but it was failed because of the huge tip-sample resistivity due to the highly insulating behavior of the CCOC samples.

In the crystal structure of CCOC, there is a single CuO$_{2}$ plane in one structural unit cell, and the CuO$_{2}$ planes are sandwiched by insulating CaCl layers. Therefore, the single crystal can be easily cleaved between the adjacent CaCl layers to expose a clean and atomically flat CaCl surface with an undistorted tetragonal crystal structure. Meanwhile, Cu atoms are directly underneath the Cl atoms on the exposed surface with a distance of about $2.75$ {\AA}. The undoped CCOC is a canonical Mott insulator, and a typical band structure of the Mott insulator is shown in Fig.~\ref{fig1}(a). For CCOC, the charge transfer gap $\Delta_\mathrm{Mott}$ is about $2.1 - 2.2$ eV from previous reports \cite{Optical,Yayu Wang1}. On the atomically flat Cl surface, we can deposit Rb atoms. The Rb adatom/cluster or the Rb monolayer grown on the surface can introduce extra electrons to the sample, which may diminish the Mott gap of the sample.

Before the Rb deposition, we characterize the CCOC substrate by doing the topographic image and tunneling spectrum measurement. An example is shown in Fig.~\ref{fig1}(c). The inset in Fig.~\ref{fig1}(c) shows the atomically resolved surface measured on CCOC sample, and one can see a perfect square lattice of Cl atoms with the lattice constant $a_{0}$  of about $3.86$ {\AA}. Here, the Cu atoms in the CuO$_{2}$ plane are just below the Cl atoms on the surface, while the O atoms are below the center between the two nearest neighboring Cl atoms. Tunneling spectra are measured along the Cl-Cl (or the Cu-O-Cu chain) direction, and they are shown in Fig.~\ref{fig1}(c). Since the CaCl layers are insulating, the differential conductance d\emph{I}/d\emph{V}, which is proportional to the local density of states (DOS), should be dominated by electronic structure in the CuO$_{2}$ plane. The spectra in Fig.~\ref{fig1}(c) show spatial homogeneity in the present scale at and between Cl sites, and there is a large gap across the Fermi energy. Features on the spectra can be understood by the schematic band structure of the Mott insulator shown in Fig.~\ref{fig1}(a). The endpoints of the gapped feature at negative (marked by green triangles in Fig.~\ref{fig1}(c)) and positive (marked by blue triangles in Fig.~\ref{fig1}(c)) energy sides correspond to the top edge of the CTB and the bottom edge of the UHB, respectively.

Then the charge transfer gap $\Delta_\mathrm{Mott}$, defined as the energy difference between CTB and UHB, is about $2.3$ eV from our data. This value is close to those reported previously by the optical reflectivity \cite{Optical} and STS \cite{Yayu Wang1} measurements. In addition, the energy of the top edge of CTB is about $-0.4$ eV which is also similar to the values obtained by ARPES measurements \cite{Zhixun Shen2,Xingjiang Zhou2}. Beyond the edges of the charge transfer gap, there are shoulders on the spectra at about $-1.0$ and $2.5$ eV and they are indicated by purple and orange arrows in Fig.~\ref{fig1}(c). It should be noted that these shoulders can exhibit as peaks on some spectra, and examples can be seen in Fig.~\ref{fig1}(d) and \ref{fig1}(e). These shoulders or peaks of local DOS at negative and positive energies may correspond to peak positions of the CTB \cite{Sawatzky2} and UHB \cite{Yayu Wang1}, respectively. The peak of CTB corresponding to prominent van Hove singularity (vHS) \cite{Jorgensen} derived from the Cu $d_{x^2-y^2}$ $-$ O $p_{x,y}$ $\sigma$-band is merged to the high energy DOS. The intense DOS appearing below $-1.5$ eV may originate from the O $2 p_{\pi}$ band \cite{Sawatzky2}. 

\begin{figure}
\includegraphics[width=8.5cm]{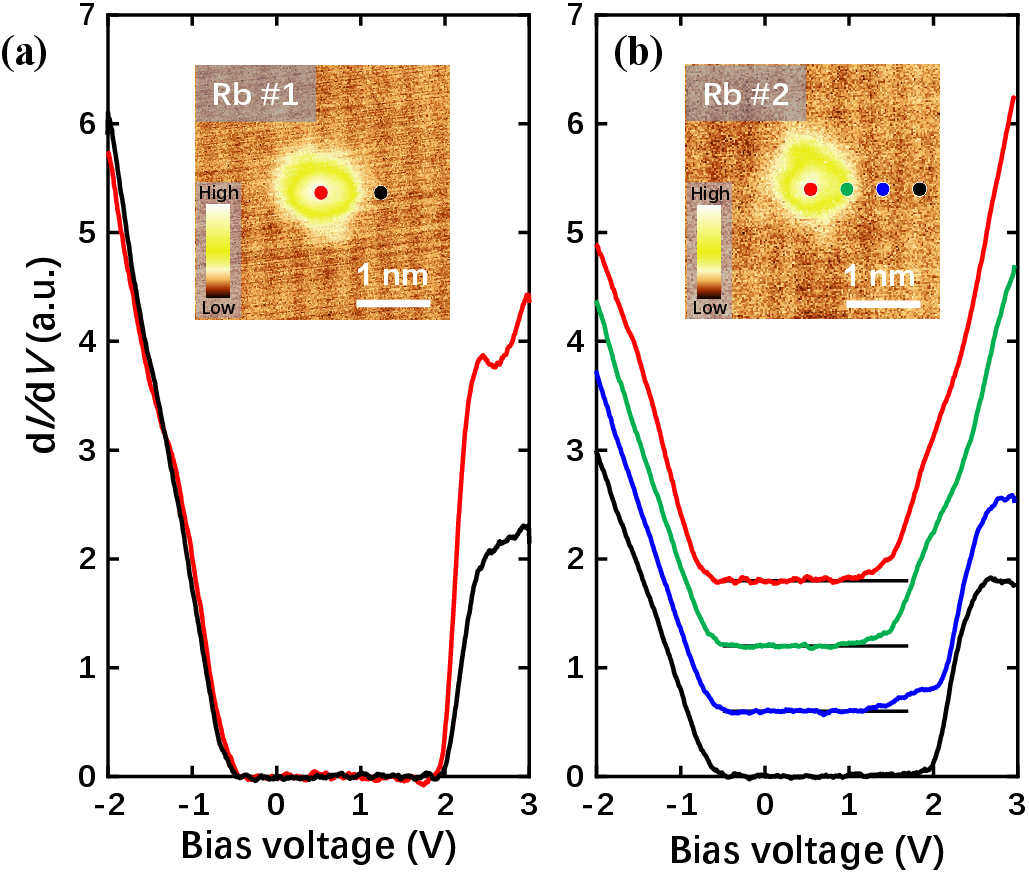}
\caption{(a) Tunneling spectra measured on and away from the Rb adatom or cluster $\#1$ deposited on CCOC surface ($V_\mathrm{set} = -2.5$ V, $I_\mathrm{set} = 100$ pA).(b) Tunneling spectra measured at different sites across the Rb atom or cluster $\#2$ ($V_\mathrm{set} = -2.5$ V, $I_\mathrm{set} = 100$ pA). Each spectrum in this figure is plotted using the same color as the point marking the measured position. Tunneling spectra in (b) are shifted for clarity.} \label{fig2}
\end{figure}

Besides the perfect Cl lattice on the surface of CCOC, we observe two types of defects: one is the Ca vacancy (Fig.~\ref{fig1}(d)), and the other is the Cl vacancy (Fig.~\ref{fig1}(e)). A Ca defect may result in the atomic relaxation of the four nearest Cl atoms on the surface and make them higher \cite{I. Tanaka}. Here, a Ca vacancy is equivalent to doping two holes to the CuO$_{2}$ plane, while a Cl vacancy donates an electron. These defects provide natural platforms for studying the Mott gap filling to the parent compound, but the defect induced DOS are quite localized. From the tunneling spectra shown in Fig.~\ref{fig1}(d) and \ref{fig1}(e), one can see that the spectra measured just above the Ca or Cl defects show similar behavior, i.e., a broad electronic state appears near the bottom edge of the UHB within the charge transfer gap. The in-gap states finally disappear on the spectrum measured $5 a_{0}$ away from the vacancy sites. Based on the theoretical calculation, hole doping may induce the in-gap state near the CTB \cite{Sawatzky1}, and the in-gap state may move to the UHB due to a repulsive impurity potential \cite{Jianxin Li1}. Meanwhile, electron doping can also induce an in-gap state below the UHB \cite{Jianxin Li2}. However, we would argue that the location of the local DOS peak induced by the Ca or Cl vacancies seem to be an effect strongly linked with the edge of the UHB, but not the repulsive impurity potential effect, since their locations and features look quite similar, despite the impurities of Cl and Ca vacancies are very different. Although the in-gap states are very similar in situations of both hole and electron doping sides, the influence on the DOS of the UHB by different kinds of doping is very different. One can see that at the Cl vacancy, the DOS show an obvious enhancement above $2$ eV, which is weaker for the Ca site.

In order to investigate the doping effect with a higher doping level than a Cl vacancy, we deposit Rb atoms to the top Cl surface of CCOC. When the depositing time is short, isolated Rb adatoms or clusters can be observed on the top surface of the Cl lattice. Insets in Fig.~\ref{fig2} show topographic images of the Rb adatom or the Rb cluster on the surface, and their diameters are both about $1$ nm while their heights are about $0.13$ and $0.10$ nm for $\#1$ and $\#2$, respectively. From the spectra shown in Fig.~\ref{fig2}(a), we did not see the peak of local in-gap DOS as formed for the Cl vacancies, which may be induced by the lower electron doping level by the Rb adatom/cluster $\#1$ than that by the Cl vacancy. The obvious effect of the Rb atom is the significant enhancement of the DOS of the UHB, but the in-gap states cannot be seen within the charge transfer gap on the spectra. For the Rb adatom/cluster $\#2$, the electron doping level seems to be higher than the Cl vacancy. The spectrum measured about 1 nm away from the adatom/cluster as plotted in blue is similar to that measured just on the Cl vacancy. When the tip moves to the positions just above the Rb adatom/cluster $\#2$, the spectral weight of the UHB is significantly enhanced. The in-gap states are also enhanced near the edge of the UHB leading to a general elevating of the DOS smoothly connected to the enhanced DOS of the UHB, meanwhile the Mott gap has also been significantly suppressed, i.e., it changes from about $2.35$ to about $1.6$ eV. The difference between the Rb adatom/cluster $\#1$ and $\#2$ may be due to the effect of different effective work functions \cite{work function}, and then the effective doping level of electrons to the CuO$_{2}$ plane. With the increase of the electronic doping level, the effective range of the Rb adatom/cluster becomes larger for $\#2$ than for $\#1$.

\begin{figure}
\includegraphics[width=8.5cm]{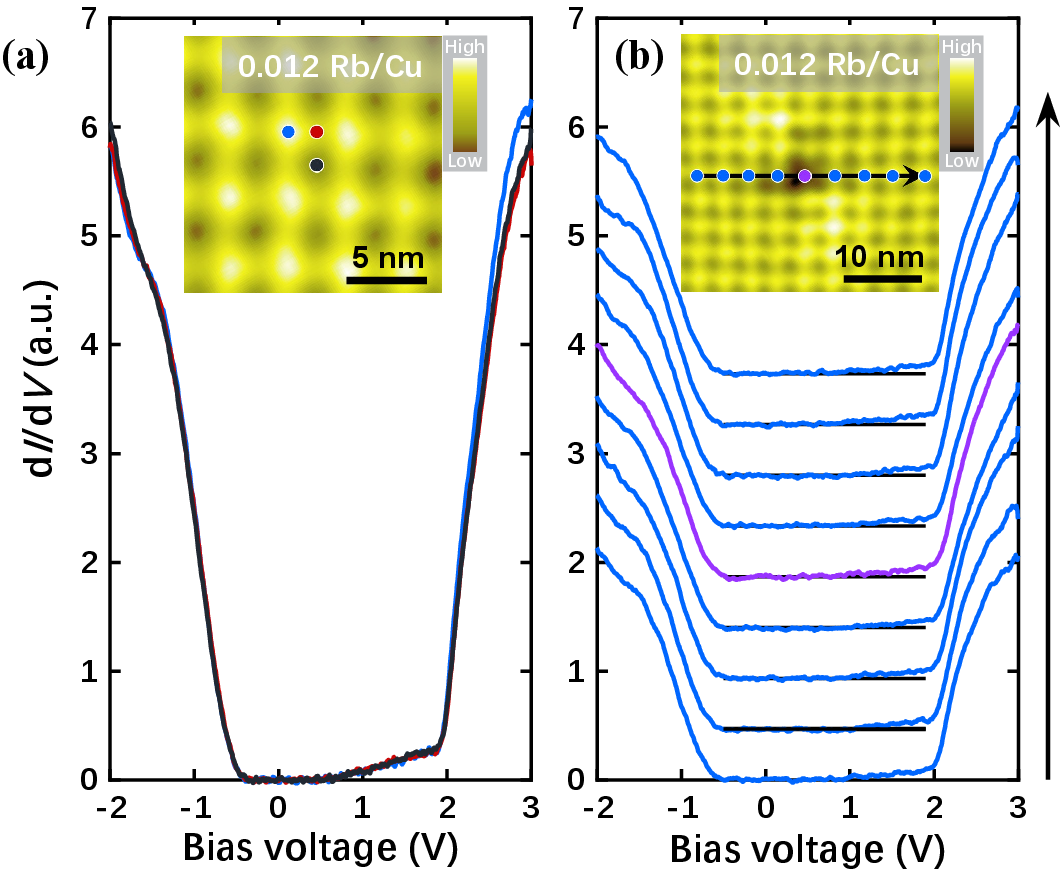}
\caption{ (a) Averaged tunneling spectra measured at different sites on the Rb film grown on the CCOC surface with the areal density of about $0.012$ Rb/Cu. The topographic image of the Rb film is shown in the inset. The typical sites are marked by colored points in the inset, and each spectrum in the main panel is plotted using the same color as the point. (b) Tunneling spectra measured across a Rb defect in the Rb film ($V_\mathrm{set} = -2.5$ V, $I_\mathrm{set} = 100$ pA). Spectra are shifted for clarity in this figure. Spectra show spatial uniformity when across the defect.}
\label{fig3}
\end{figure}

With longer depositing time, Rb atoms form an ordered lattice on the CCOC surface instead of the isolated Rb adatoms/clusters. The inset in Fig.~\ref{fig3}(a) shows a topographic image of a Rb film with an areal density of about $0.012$ Rb/Cl (or Rb/Cu). This is a rectangular lattice with lattice constants of about $35$ and $38$ {\AA} with a small anisotropy of about $1.08$. The dosed atoms of alkali metal usually have a very disordered distribution \cite{Q. K. Xue1,Q. K. Xue2} or form some locally ordered lattice \cite{Donglai Feng1,M. G. Betti} on different substrates, as far as we know the ordered lattice in a very large scale has not been observed before. Here it should be noted that the Rb atoms are grown on the Mott insulator of CCOC, and the valence electrons of Rb atoms are supposed to go to the CuO$_{2}$ plane, leaving only Rb$^{1+}$ irons on the top layer. The Cl irons on the top surface of the substrate form a perfect structure with one block unit layer of the copper oxide octahedra containing a single CuO$_{2}$ plane,thus they should have closed shells of electrons and are insulating. Therefore, the interaction between Rb irons and the substrate may be negligible, and the repulsive force between Rb irons makes them to form a very ordered lattice. In this point of view, the measurements on the lattice of Rb irons on the surface can reflect the Mott insulating nature of the CuO$_{2}$ plane of the substrate of CCOC.

Since we have successfully grown the Rb film with a regular lattice on CCOC, it is very interesting to investigate the electronic properties of the sample. Strikingly, here the size of the Rb iron determined from the topography seems to be quite large, i.e., the effective diameter of a Rb iron seems to be close to the lattice constant of ($3.6$ nm on average). This is certainly not the real size, but showing an extended influence of DOS affected by the Rb ions. Then we measure the tunneling spectra at the Rb centers, in the middles of the two nearest neighbored Rb ions, and at the centers of a rectangle block of four Rb ions, and we plot the averaged spectra measured at these three positions in Fig.~\ref{fig3}(a). One can see that the spectra almost overlap with each other except for the part in the high energy region for the UHB. The DOS is slightly enhanced above $2$ eV at the Rb site, which may suggest a slightly higher electron doping level at the center of the Rb ion. However, the in-gap states have very little energy dependence at different positions. In some places, we can see Rb vacancies, and one example is shown in the inset of Fig.~\ref{fig3}(b). However, based on the spectra shown in Fig.~\ref{fig3}(b), the Rb vacancy has negligible influence on the spectrum. In other words, the neighboring Rb ions to the vacancy has an influential range of about $3.6$ nm which is much larger than that of $3 a_{0} \approx 1.2$ nm estimated for the Ca or Cl vacancies, as shown in Fig.~\ref{fig1}. Meanwhile, it should be considered that the Rb vacancy is surrounded by Rb ions, as we argued above, this will give rise to extended DOS to the vacancy site.

\begin{figure}
\includegraphics[width=8.5cm]{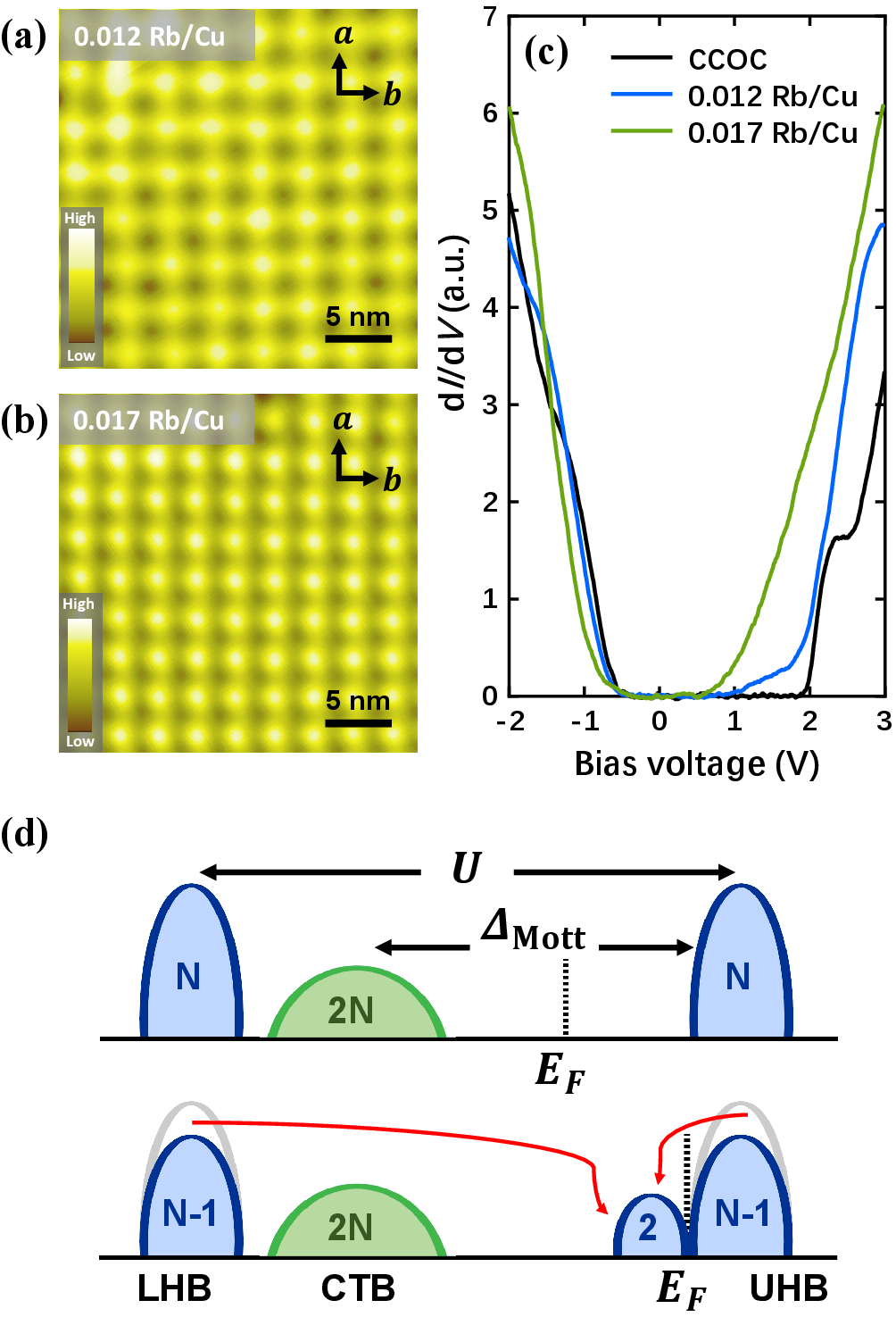}
\caption{Topographic image of Rb films grown on the CCOC surface with the areal density of (a) about $0.012$ Rb/Cu and (b) about $0.017$ Rb/Cu ($V_\mathrm{set} = -2.5$ V, $I_\mathrm{set} = 100$ pA). (c) Averaged tunneling spectra measured on CCOC and Rb films with different areal density ($V_\mathrm{set} = -2.5$ V, $I_\mathrm{set} = 100$ pA). (d) A schematic drawing of the electron-addition DOS for the insulating undoped charge-transfer model.
} \label{fig4}
\end{figure}

In the Rb monolayer grown on the CCOC substrate, the electron states are almost homogeneous in nanoscale. With the increase of the deposition time, the Rb monolayer with different areal densities can be obtained. Fig.~\ref{fig4}(a) and \ref{fig4}(b) display two examples of the Rb monolayer with different lattice constants. The lattice shown in Fig.~\ref{fig4}(a) is the same as those shown in Fig.~\ref{fig3}, and the areal density is about $0.012$ Rb/Cu. The rectangular lattice has the lattice constants of $3.8$ and $3.5$ nm. Fig.~\ref{fig4}(b) shows the topography of a Rb monolayer with an areal density of about $0.017$ Rb/Cu, and the lattice now becomes almost a square shape with a lattice constant of about $2.9$ nm. The change from the rectangular lattice to the standard square lattice may be induced by a larger areal density of Rb resulting in a stronger repulsive interaction between two neighbored Rb irons. Fig.~\ref{fig4}(c) shows the averaged tunneling spectra measured on different Rb monolayers. One can see that with the increase of the areal density of Rb, the in-gap states are significantly enhanced and then overlapped with the simultaneously enhanced UHB, resulting in a clear suppression of the Mott gap.


From our results, a striking observation is that the chemical potential of Cl-vacancy, Ca-vacancy, Rb adatoms or clusters and Rb films all stays fixed in the charge transfer gap but is closer to the CTB, which echoes with the pinned chemical potential at low doping \cite{R. Liu,S. Uchida}. This is different from the picture with a chemical potential shift or jump observed in other p-type and n-type cuprates \cite{Xingjiang zhou1,Ref2A,Ref2B,Ref2C} . This may be explained by the small doping level of the Rb film. However, the chemical potential also stays fixed in the charge transfer gap among Ca and Cl vacancies, this is contrast to the expectation of the rigid model for a doped Mott system \cite{Sawatzky1}.

The extraordinary large lattice constant of Rb films have never been found on cuprate system before. The lattice constant in $a$ and $b$ directions for $0.012$ Rb/Cu film are $\sim$$10$ and $\sim$$9$ times the lattice constants of CCOC, and both $\sim$$7.6$ times for $0.017$ Rb/Cu film. The radii of the isolated Rb atoms and corresponding ion are $2.48$ {\AA} and $1.52$ {\AA}, respectively \cite{D. J. Shu}. The spatial scale of Rb ions for $0.012$ and $0.017$ Rb/Cu film defined by full width at half maximum are $\sim$$21$ {\AA} and $\sim$$17$ {\AA}, respectively. It is $\sim$$14$ times the diameter of Rb cation for $0.012$ Rb/Cu film and $\sim$$11$ times for $0.017$ Rb/Cu film. As mentioned before, this size is certainly much larger than the real size of an individual atom or ion of Rb in vaccum, which is supposed to be induced by extended DOS of the Rb ion on the background of CCOC. We propose a scenario to comprehend the uniform electronic structure throughout the Rb lattice with an extraordinarily large radii of Rb ions. When the density of adsorbed Rb atoms is beyond a certain threshold, a two-dimensional Rb film with a regular lattice is formed. The donated electron by a Rb atom will be absorbed by one Cu$^{2+}$ to form a double occupancy state, which may locally enhance the hopping but weaken the correlation effect. This picture implies that the doped electrons donated by Rb atoms would dilute the spin system by neutralizing the spin on a 3 $d^9$ Cu site and adjust the hopping energy $t$ and super-exchange energy $J$, forming a circular hopping active puddle centered on the Rb donor atom. For a single Rb atom on the CCOC background, this effect extends to a large distance compared with of the radii of a Rb individual atom or ion, but with a gradually decreased strength from the site of Rb to far away. Thus, the modified electronic wave function given by an individual Rb atom looks like a puddle. Each puddle contains one double occupancy state. When many Rb atoms form a reconstructed lattice, these puddles will overlap each other in the lattice. However, the state formed by the puddles is bounded by the Rb film area, and the motion of this double occupancy state is still prohibited by the correlation effect, resulting in a nontrivial insulating state.

The picture we proposed above can also explain why the measured size of each Rb atom looks so large, that is because the double occupancy state given by the donated electron of each Rb atom may be mobile to a certain distance which is much larger than the radii of the Rb atom itself. Several interesting properties remain to be explored to explain the large lattice constants, such as binding distance, charge transfer, and work function as adsorbed Rb atoms change from clusters to two-dimensional lattice. According to the rigid band picture for doping electrons to the Mott insulator \cite{Sawatzky1}, when one electron is doped to the system, both the LHB (or CTB) and the UHB will lose one electron state, but an new electron state with two electrons will appear at the lower edge of the UHB, and the Fermi energy will immediately jump from an in-gap energy to the lower edge of the UHB, see Fig.~\ref{fig4}(d). However, our experiments do not support this picture. Firstly, the Fermi energy seems to be pinned in the charge transfer gap. Secondly, the DOS weight of the UHB gets enhanced, instead of weakened. Meanwhile, some DOS start to accumulate in the charge transfer gap region near the edge of the UHB. We believe the electron doping will systematically change the electron hopping term $t$ and the superexchange $J$ values, thus leading to a complex modification of the whole electronic state. Our experiments expose distinctions from the rigid band model, which sheds new light in understanding the evolution of the doped Mott insulator, and intimately helps unraveling the mystery of the superconductivity mechanism in cuprates.

We have successfully grown thin films of Rb with periodic structures on CCOC parent compound, which allows a detailed investigation on the doped Mott insulator of the parent phase of cuprate. The STM measurements of the Rb film revealed that the Fermi energy is close to the upper edge of the charge transfer band and is pinned within the Mott gap. We also find that the spectral weight of UHB is obviously enhanced by electrons from Rb donor atoms, in stark contrast to the expectation of the rigid band model which postulates that the weight of the UHB will be suppressed after doping electrons. Meanwhile, with increasing electron doping, broad in-gap states gradually set up near the bottom of UHB, and the Mott gap will be significantly diminished. The spectra measured on Rb thin films and Rb atoms or clusters show similar behavior. Our work provides a new clue to the understanding of Mott physics and sheds new light on the origin of high-temperature superconductivity.

The work was supported by the National Natural Science Foundation of China (Grants No. 11974171, No. 12061131001, and No. 11927809) and the National Key R\&D Program of China (Grant No. 2022YFA1403201).

$^*$ huanyang@nju.edu.cn 

$^\dag$ hhwen@nju.edu.cn


\begin{thebibliography}{40}

\bibitem{Xiaogang Wen} P. A. Lee, N. Nagaosa, and X.-G. Wen, Rev. Mod. Phys. \textbf{78}, 17 (2006).

\bibitem{Kivelson} B. Keimer, S. A. Kivelson, M. R. Norman, S. Uchida, and J. Zaanen, Nature (London) \textbf{518}, 179 (2015).

\bibitem{Anderson 1959} P. W. Anderson, Phys. Rev. \textbf{115}, 2 (1959).

\bibitem{RBV 1987} P. W. Anderson, Science \textbf{235}, 1196 (1987).

\bibitem{J. E. Hirsch} J. E. Hirsch, Phys. Rev. B \textbf{31}, 4403 (1985).

\bibitem{t-J Model} K. A. Chao, J. Spałek, and A. M. Oleś, Phys. Rev. B \textbf{18}, 3453 (1978).

\bibitem{J. Zaanen1} J. Zaanen, G. A. Sawatzky, and J. W. Allen, Phys. Rev. Lett. \textbf{55}, 418 (1985).

\bibitem{p-d model} V. J. Emery, Phys. Rev. Lett. \textbf{58}, 2794 (1987).

\bibitem{AFM} P. W. Anderson, Phys. Rev. \textbf{79}, 350 (1950).

\bibitem{Stripe}  J. M. Tranquada, B. J. Sternlieb, J. D. Axe, Y. Nakamura and S. Uchida, Nature (London) \textbf{375}, 561 (1995).

\bibitem{CDW} G. Ghiringhelli, M. Le Tacon, M. Minola, S. Blanco-Canosa, C. Mazzoli, N. B. Brookes, G. M. De Luca, A. Frano, D. G. Hawthorn, F. He, T. Loew, M. Moretti Sala, D. C. Peets, M. Salluzzo, E. Schierle, R. Sutarto, G. A. Sawatzky, E. Weschke, B.Keimer and L. Braicovich, Science \textbf{337}, 821 (2012).

\bibitem{Sawatzky1} H. Eskes, M. B. J. Meinders, and G. A. Sawatzky, Phys. Rev. Lett. \textbf{67}, 1035 (1991).

\bibitem{W. A. Groen} M. A. van Veenendaal, G. A. Sawatzky, and W. A. Groen, Phys. Rev. B \textbf{49}, 1407 (1994).

\bibitem{Ref1A} N. P. Armitage, P. Fournier, and R. L. Greene, Rev. Mod. Phys. \textbf{82}, 2421 (2010).

\bibitem{Ref1B} R. L. Greene, P. R. Mandal, N. R. Poniatowski, and T. Sarkar, Annu. Rev. Condens. Matter Phys. \textbf{11}, 213 (2020).

\bibitem{Ref1C} C. W. Chu, L. Z. Deng, and B. Lv, Physica C \textbf{514}, 290 (2015).

\bibitem{Ref1D} C. Weber, K. Haule, and G. Kotliar, Nat. Phys. \textbf{6}, 574 (2010).

\bibitem{Ref1E} E. Pavarini, I. Dasgupta, T. Saha-Dasgupta, O. Jepsen, and O. K. Andersen, Phys. Rev. Lett. \textbf{87}, 047003 (2001).

\bibitem{Haihu Wen} H. Z. Li, H. Li, Z. H. Wang, S. Y. Wan, H. Yang, and H. H. Wen, npj Quantum Mater. \textbf{8}, 18 (2023).

\bibitem{Takagi1} Y. Kohsaka, M. Azuma, l. Yamada, T. Sasagawa, T. Hanaguri, M. Takano, and H. Takagi, J. Am. Chem. Soc. \textbf{124}, 12275 (2002).

\bibitem{Davis1} T. Hanaguri, C. Lupien, Y. Kohsaka, D. H. Lee, M. Azuma, M. Takano, H. Takagi, and J. C. Davis, Nature (London) \textbf{430}, 1001 (2004).

\bibitem{Takagi2}  T. Hanaguri, Y. Konsaka, J. C. Davis, C. Lupien, l. Yamada, M.  Azuma, M. Takano, K. Ohishi, M. Ono, and H. Takagi, Nat. Phys. \textbf{3}, 865 (2007).

\bibitem{Zhixun Shen1} K. M. Shen, F. Ronning, D. H. Lu, F. Baumberger, N. J. C. Ingle, W. S. Lee, W. Meevasana, Y. Kohsaka, M. Azuma, M. Takano, H. Takagi, and Z. X. Shen, Science \textbf{307}, 901 (2005).

\bibitem{Xingjiang zhou1} C. Hu, J. F. Zhao, Q. Gao, H. T. Yan, H. T. Rong, J. W. Huang, J. Liu, Y. Q. Cai, C. Li, H. Chen, L. Zhao, G. D. Liu, C. Q. Jin, Z. Y. Xu, T. Xiang, and X. J. Zhou, Nat. Commun. \textbf{12}, 1356 (2021).

\bibitem{Takano} Z. Hiroi, N. Kobayashi, and M. Takano, Nature (London) \textbf{371}, 139 (1994).

\bibitem{Optical} K. Waku, T. Katsufuji, Y. Kohsaka, T. Sasagawa, H. Takagi, H. Kishida, H. Okamoto, M. Azuma, and M. Takano, Phys. Rev. B \textbf{70}, 134501 (2004).

\bibitem{Yayu Wang1} C. Ye, P. Cai, R. Z. Yu, X. D. Zhou, W. Ruan, Q. Q. Liu, C. Q. Jin, and Y. Y. Wang, Nat. Commun. \textbf{4}, 1365 (2013).

\bibitem{Zhixun Shen2} F. Ronning, C. Kim, D. L. Feng, D. S. Marshall, A. G. Loeser, L. L. Miller, J. N. Eckstein, I. Bozovic, and Z. X. Shen, Science \textbf{282}, 2067 (1998).

\bibitem{Xingjiang Zhou2} C. Hu, J. F. Zhao, Y. Ding, J. Liu, Q. Gao, L. Zhao, G. D. Liu, L. Yu, C. Q. Jin, C. T. Chen, Z. Y. Xu, and X. J. Zhou,  Chin. Phys. Lett. \textbf{35}, 067403 (2018).

\bibitem{Sawatzky2} J. J. M. Pothuizen, R. Eder, N. T. Hien, M. Matoba, A. A. Menovsky, and G. A. Sawatzky, Phys. Rev. Lett. \textbf{78}, 717 (1997).

\bibitem{Jorgensen} D. L. Novikov, A. J. Freeman, and J. D. Jorgensen, Phys. Rev. B \textbf{51}, 6675 (1995).

\bibitem{I. Tanaka} Y. Kumagai, F. Oba, I. Yamada, M. Azuma, and I. Tanaka, Phys. Rev. B \textbf{80}, 085120 (2009).

\bibitem{Jianxin Li1} C. P. He, S. L. Yu, T. Xiang, and J. X. Li, Chin. Phys. Lett. \textbf{39}, 057401 (2022).

\bibitem{Jianxin Li2} W. H. Leong, S. L. Yu, T. Xiang, and J. X. Li, Phys. Rev. B \textbf{90}, 245102 (2014).

\bibitem{work function} J. P. Muscat and D. M. Newns, J. Phys. C: Solid St. Phys, \textbf{7}, 2630 (1974).

\bibitem{Q. K. Xue1} C. L. Song, H. M. Zhang, Y. Zhong, X. P. Hu, S. H. Ji, L. Wang, K. He, X. C. Ma, and Q. K. Xue, Phys. Rev. Lett. \textbf{116}, 157001 (2016).

\bibitem{Q. K. Xue2} Y. H. Yuan, X. M. Fan, X. T. Wang, K. He, Y. Zhang, Q. K. Xue, and W. Li, Nat. Commun. \textbf{12}, 2196 (2021).

\bibitem{Donglai Feng1} M. Q. Ren, Y. J. Yan, X. H. Niu, R. Tao, D. Hu, R. Peng, B. P. Xie, J. Zhao, T. Zhang, and D. L. Feng, Sci. Adv. \textbf{3}, e1603238 (2017).

\bibitem{M. G. Betti} L. Gavioli, M. Padovani, E. Spiller, M. Sancrotti and M. G. Betti, Appl. Surf. Sci. \textbf{212}, 47 (2003).

\bibitem{R. Liu} J. W. Allen, C. G. Olson, M. B. Maple, J. S. Kang, L. Z. Liu, J. H. Park, R. O. Anderson, W. P. Ellis, J. T. Markert, Y. Dalichaouch, and R. Liu, Phys. Rev. Lett. \textbf{64}, 595 (1990).

\bibitem{S. Uchida} A. Ino, C. Kim, M. Nakamura, T. Yoshida, T. Mizokawa, Z. X. Shen, A. Fujimori, T. Kakeshita, H. Eisaki, and S. Uchida, Phys. Rev. B \textbf{62}, 413 (2000).

\bibitem{Ref2A} N. Harima, J. Matsuno, A. Fujimori, Y. Onose, Y. Taguchi, and Y. Tokura, Phys. Rev. B \textbf{64}, 220507(R) (2001).

\bibitem{Ref2B} N. P. Armitage, F. Ronning, D. H. Lu, C. Kim, A. Damascelli, K. M. Shen, D. L. Feng, H. Eisaki, Z.-X. Shen, P. K. Mang, N. Kaneko, M. Greven, Y. Onose, Y. Taguchi, and Y. Tokura, Phys. Rev. Lett. \textbf{88}, 257001 (2002).

\bibitem{Ref2C} M. Taguchi, A. Chainani, K. Horiba, Y. Takata, M. Yabashi, K. Tamasaku, Y. Nishino, D. Miwa, T. Ishikawa, T. Takeuchi, K. Yamamoto, M. Matsunami, S. Shin, T. Yokoya, E. Ikenaga, K. Kobayashi, T. Mochiku, K. Hirata, J. Hori, K. Ishii, F. Nakamura, and T. Suzuki, Phys. Rev. Lett. \textbf{95}, 177002 (2005).

\bibitem{D. J. Shu} Z. Y. Wang, and D. J. Shu, J. Phys. Chem. C \textbf{125}, 19259 (2021).

\end{thebibliography}
\end{document}